\documentclass{svproc}
\usepackage[utf8]{inputenc}
\usepackage{graphicx}
\usepackage{enumitem}
\usepackage{amsmath}

\title{A methodology for analyzing financial needs hierarchy from social discussions using LLM}
\titlerunning{Methodology to analyze financial needs hierarchy from social discussions}

\author{Abhishek Jangra  \and Sachin Thukral \and  Arnab Chatterjee  \and Jayasree Raveendran}

\authorrunning{A. Jangra et al.}

\institute{TCS Research\\
\email{abhishek.jangra, sachi.2, arnab.chatterjee4, jayasree.raveendran@tcs.com}}

\begin{document}

\maketitle

\begin{abstract}
This study examines the hierarchical structure of financial needs as articulated in social media discourse, employing generative AI techniques to analyze large-scale textual data. While human needs encompass a broad spectrum—from fundamental survival to psychological fulfillment—financial needs are particularly critical, influencing both individual well-being and day-to-day decision-making. Our research advances the understanding of financial behavior by utilizing large language models (LLMs) to extract and analyze expressions of financial needs from social media posts. We hypothesize that financial needs are organized hierarchically, progressing from short-term essentials to long-term aspirations, consistent with theoretical frameworks established in the behavioral sciences. Through computational analysis, we demonstrate the feasibility of identifying these needs and validate the presence of a hierarchical structure within them. In addition to confirming this structure, our findings provide novel insights into the content and themes of financial discussions online. By inferring underlying needs from naturally occurring language, this approach offers a scalable and data-driven alternative to conventional survey methodologies, enabling a more dynamic and nuanced understanding of financial behavior in real-world contexts.


\end{abstract}

\section{Introduction}

The rise of AI and social media offers an exciting opportunity to analyze real-world behaviors. Needs are origins to human behaviors and financial needs are a subset of overarching needs.
With abundant data on financial discussions across online platforms, AI can help extract insights into individuals' financial needs.

Human needs range from basic survival to higher-order psychological goals, as conceptualized by frameworks like Maslow’s Hierarchy~\cite{maslow1943theory}. Financial needs play a central role in this spectrum, supporting both immediate survival and overall well-being. Prior research shows that financial motivations, such as spending and saving, align closely with broader human needs~\cite{xiao1994perceived}.

Although needs hierarchies have been studied for decades, empirical tests for a hierarchy of financial needs are rare. Analyzing needs expressed on social platforms offers a novel way to validate this structure. This study proposes a method to extract financial needs from social data and tests the hierarchy using age and income information, contributing new insights to the literature.

Despite research on human needs, empirical evidence for a hierarchy of financial needs at the individual level is limited. While studies have examined saving behaviors~\cite{lee2015savings}, money attitudes~\cite{oleson2004exploring}, and financial well-being~\cite{kempson2017financial}, few have explored how financial needs fit within broader psychological hierarchies. Prior work has addressed household financial needs~\cite{xiao1994perceived,xiao1997hierarchical} and individual savings motivation~\cite{canova2005hierarchical}, but generalizable insights remain scarce. Understanding hierarchies is crucial for analyzing user behavior, as it provides insight into how individuals prioritize their financial concerns and make decisions based on their immediate versus future aspirations. 

This study examines financial needs expressed in social media discussions. We propose a method to extract and validate a financial needs hierarchy, hypothesizing a progression from short-term essentials to long-term goals. Drawing on Maslow’s hierarchy~\cite{maslow1943theory} and García-Mata \& Zerón-Félix’s framework~\cite{garcia2022review}, we find evidence supporting this structure. Ref~\cite{maslow1943theory} categorizes human needs into five levels, starting with basic survival needs at the bottom of the hierarchy. Once these are met, individuals focus on safety (security, stability), love and belonging (family, relationships), esteem (status, prestige), and finally self-actualization. Ref~\cite{garcia2022review} proposed a financial need prioritization framework having three levels, namely, consumption for meeting everyday needs, savings to handle emergencies, retirement savings, wealth, and lifestyle improvement.

The key contributions of this paper are:
\begin{enumerate}[label=(\roman*)]
    \item We introduce a methodology to explore the needs hierarchy theorized in two major frameworks, viz. NHF and NPF.
    \item Our methodology employs state of the art techniques from Natural Language Processing (NLP) and Generative AI (Gen-AI) to extract information about the needs and their hierarchies, along with various behavioral attributes for users from a social discussion dataset. 
    \item This helps in comparing and correlating the needs information with various demographic attributes in a unique method vis-à-vis the traditional methods used in empirical finance literature, which heavily depends on panel data created out of survey instruments.
\end{enumerate}

The objective of the paper is to bring empirical evidence to the financial needs hierarchy. This is done by analyzing the needs expressed as queries in the social media discussions. We add evidence to confirming the known behavioral patterns of how needs manifest in the discussions through a novel method to extract and identify the hierarchy. 

In the following, Section~\ref{sec:relatedwork} presents the literature review, Section~\ref{sec:hypo} discusses hypotheses development, Section~\ref{sec:method} elaborates the methodology adopted, Section~\ref{sec:results} discusses the analysis and findings, followed by Discussions. 

\section{Related work}
\label{sec:relatedwork}

Financial needs, as a subset of human needs, shape both present spending and future planning. Maslow’s hierarchy~\cite{maslow1943theory} describes motivations from survival to self-actualization, while alternative models like Alderfer’s three-level~\cite{alderfer1969empirical} and Lindqvist’s four-level savings motives~\cite{lindqvist1981note} highlight priorities such as security and wealth. Research shows families often meet basic needs before pursuing aspirational goals~\cite{xiao1994perceived,kitces2010emerging,garcia2022review}, though some frameworks argue needs progress linearly~\cite{maslow1954motivation} while others suggest they can coexist~\cite{alderfer1969empirical}. Overall, comprehensive studies on the hierarchy of financial needs remain limited~\cite{xiao1994perceived,canova2005hierarchical}.

Social media platforms capture complex societal behaviors, with prior research showing their potential in predicting market variables like returns and volatility~\cite{oliveira2017impact}. NLP techniques, such as topic modeling, have also provided insights into financial discussions on platforms like Reddit~\cite{karpenko2021study}.

The rise of Large Language Models (LLMs) has enhanced the ability to analyze financial sentiment and behavior, e.g., develop market sentiment models~\cite{deng2023llms} and to serve as personal financial advisors for optimized decision-making~\cite{lakkaraju2023can}. Furthermore,~\cite{de2023optimized} proposed an LLM-driven framework that integrates individual and cooperative budgeting models to improve financial planning.

Financial needs are shaped by behavioral factors such as stress, risk propensity, and emotions. Unmet basic needs increase anxiety and financial stress, especially when societal pressure drives spending on higher-order needs beyond one’s means~\cite{northern2010development,coate1988human,conger2007interactionist,lusardi2011financial,forgue1996personal}. Individuals with unmet lower-order needs tend to avoid risk, while those pursuing higher-order goals take more risks as income grows~\cite{bricker2021signaling,roussanov2010diversification,weagley1991investor}. Emotions like fear, happiness, and confidence also influence financial choices~\cite{shim2009pathways}. This study uses social media data to examine financial stress, risk propensity, and emotion.

\section{Hypotheses}
\label{sec:hypo}

We hypothesize that individual financial needs follow a hierarchy, from short-term necessities to long-term aspirations. This is tested using the NHF~\cite{maslow1943theory} and the NPF~\cite{garcia2022review}. Based on existing literature and identified gaps, we propose hypotheses regarding financial needs, hierarchy, prioritization, and income.

\begin{itemize}
    \item [H1:]  Individual financial needs follow a hierarchical structure and can be mapped using NHF~\cite{maslow1943theory}. 
There is a positive correlation between income and hierarchy of financial needs within the NHF. 
\item [H2:]
Individual financial needs follow a priority order and can be mapped using the NPF~\cite{garcia2022review}. 
There is a positive correlation between income and the priority of financial needs within the NPF.
\end{itemize}

\section{Methodology}
\label{sec:method}

\begin{figure}[t]
    \centering
    \includegraphics[width=0.8\linewidth]{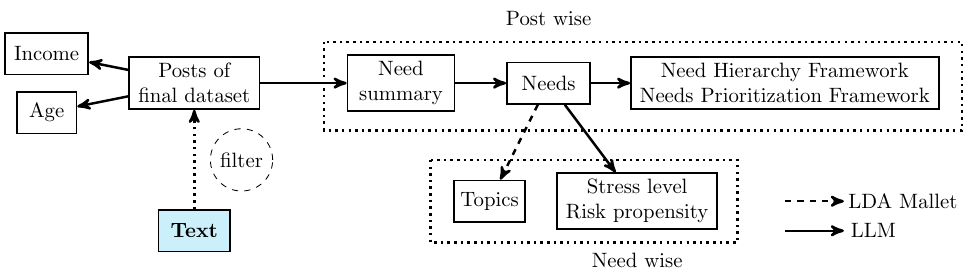}
    \caption{Framework for extraction of various features from user text data}
    \label{fig:pipeline}
\end{figure}

We use a novel, data-driven approach with LLMs to analyze public financial posts on Reddit related to personal finance. This study combines qualitative methods (extracting text data with LLMs) and quantitative analysis for statistical insights, helping us map financial needs to hierarchical frameworks and better understand individual financial behavior from social media data.
We developed a framework (Figure~\ref{fig:pipeline}) to process and analyze user posts, which consists of the following steps: 
\begin{enumerate}[label=(\roman*)]
    \item First of all, we use a series of steps to filter out the redundant posts as well to increase the chances of capturing information regarding age and income from the text. Users with at least 15 posts, and with explicit mention of age and income in at least one of the posts are selected. Posts with 20 words or more are considered. 
    \item We create summaries of each post using LLM prompts, to identify core queries and additional queries. 
    \item We extract needs from the created summaries using LLM prompts.
    \item From the needs, 
    we map the needs to the respective hierarchies based on two frameworks, as well as income, using LLM prompts. 
\item 
    We extract three behavioral attributes – risk propensity and stress level at the level of needs using LLM prompts, and emotions at the post level. 
\item 
    We extract topics from needs using LDA MALLET. 
\end{enumerate}

Detailed descriptions of the above steps and their rationale are given in the feature extraction section (Section~\ref{sec:features}). 

\subsection{Data description}
Reddit is a social network for topical discussions, organized into subreddits where users create posts on issues of their interest, to discuss and engage with others, sharing information, asking questions to solve their issues or simply engage with others who share similar interests.
We collected data using the Pushshift API~\cite{Reddit_dump} from four subreddits - \texttt{r/personalfinance}, \texttt{r/FinancialPlanning}, \texttt{r/investing}, and \texttt{r/EstatePlanning}, covering posts from January 1, 2020, to December 31, 2023. Our dataset includes 614,089 posts in \texttt{r/personalfinance}, 48,554 in \texttt{r/FinancialPlanning}, 184,382 in \texttt{r/investing}, and 4,454 in \texttt{r/Estate Planning}. We focused on 13,821 posts made by 641 distinct users, considering only those who mentioned their age and income. The final dataset consists of 6,709 posts from 334 users, averaging 21 posts per user.
Every post contains post id, username, timestamp and the text written by the user. Although each post may be followed by a comment thread contributed by various users (where the discussions take place), we limit our analysis only to the posts, in this study.

Our analysis specifically focuses on posts made by Reddit users, as needs are primarily expressed in the form of queries. In future work, we may extend this to include comments to gain additional insights into how these needs are addressed or fulfilled by the community.

\subsection{Feature Extraction}
\label{sec:features}

In this study, we extracted various features from each post, including personal information like (i) age, (ii) income, (iii) financial needs with hierarchy mapping, behavioral features like (iv) emotion, (v) financial stress, (vi) risk propensity, and (vii) discussion topics from needs. 
We extracted these features to understand users' financial situations and behaviors as reflected in social media discussions. Personal details like age, income, and financial needs provide context, while behavioral cues such as emotion, stress, and risk propensity reveal underlying psychological phenomena. Discussion topics further highlight user priorities and concerns.
We used LLMs from the Llama family~\cite{dubey2024llama3herdmodels} via the Groq API~\footnote{Groq API \texttt{https://groq.com/groqcloud/}}to extract most features, except for topics, which were derived using the LDA MALLET toolkit~\cite{blei2003latent}. Prompts were fine-tuned through manual validation on a subset for accuracy. The extraction processes for these features are as follows:

\noindent \textbf{Age:}
We designed LLM prompts to extract exact age and age category (Less than 21, 21-30, 31-40, 41-50, 51-60, More than 60). If age is not explicitly mentioned, the output is ``not mentioned." Since users may not mention their age in every post, we map the user's age to all available posts using the following method:
\begin{enumerate}[label=(\roman*)]
    \item We use the most recent mention of the user's age. For example, if age is mentioned in both 2022 and 2023, we take the age from 2023. 
    \item
    If a user's age is mentioned multiple times in a year (e.g., 2023), we take the lowest age to avoid overestimating their age and misrepresenting their financial stage. For example, if a user mentions being 33 and 35, we use 33 for all posts in that year.
\end{enumerate}
\noindent
\textbf{Income}
Income refers to the monthly cash inflow from salary and other sources in the current financial year. Any non-monthly income mentions are converted to monthly values. We found that all income data was reported in US Dollars, indicating that the users were geographically localized in the USA. Since users may not mention income in every post, we map their reported income to all posts during the sample period.
\begin{enumerate}[label=(\roman*)]
    \item If income is not mentioned, all the user's posts are excluded from analysis.  
\item If income is mentioned once in a year, it is mapped to all posts in that year.  
\item  If income is mentioned multiple times in a year, the lowest value is  mapped to all posts that year, as well as to posts from preceding years with no income mentioned.
\end{enumerate}

\noindent
\textbf{Financial Needs}
Financial needs are extracted for each post using the following method: 
\begin{itemize}
    \item[(i)] Using LLM, we create one core query and two additional queries from the financial post. 
    \item[(ii)] Subsequently, we extract financial needs in terms of need label which is a combination of purpose (medical expenses, education, buying house, etc.) and financial process (saving, investing, budgeting, insurance, etc.).
 \end{itemize}
 
\noindent
\textbf{Financial Needs Hierarchy}
The financial needs are mapped to hierarchies using two frameworks: (i) \textit{needs hierarchy framework (NHF)}, and (ii) \textit{financial needs prioritization framework (NPF)}.  
For NHF mapping, we consider 7 levels, namely: 
\begin{itemize}
    \item basic needs (Rent and utilities, Debt management strategies),
    \item  safety needs 
    \begin{itemize}
        \item level 1 (Emergency savings, Basic insurance coverage),
        \item level 2 (Purchasing a home, Long-term savings for retirement),
    \end{itemize} 
    \item love and belongingness (Family gatherings, Support for aging parents), 
    \item esteem needs (Investment strategies in real estate, Achieving financial independence through self-employment),
    \item self-transcendence (Financial freedom of choices, Exploring passions and opportunities),
    \item self-actualization (Philanthropy and charitable contributions, Legacy planning).
\end{itemize} 
However, for further analysis, we combine safety needs - level 1 and safety needs - level 2 as Safety Needs, and self-actualization and self-transcendence as self-actualization, finally arriving at 5 categories.

In NPF, we use 3 categories: (i) Consumption and Immediate Needs, (ii) Savings for Emergency Provisions, and (iii) Retirement Savings, Wealth, and Lifestyle Improvement. Emergency provisions include saving for emergencies and stability needs. The ``Retirement Savings, Wealth, and Lifestyle Improvement" category includes sub-categories like long-term investments, wealth creation, and lifestyle improvement, which covers real estate, vacations, charitable donations, and hobbies. For insights, we focus on the original three categories.

\subsection{Topic Extraction}
\label{sec:topicex}

Topic extraction identifies key themes from text, providing a high-level overview of the content in large repositories using unsupervised models, offering insights into individuals' financial priorities, challenges, and behaviors. In our work, we are using established approach for extracting topics and determining the optimal number of topics ($k$) from financial needs. We utilize the Latent Dirichlet Allocation (LDA) MALLET toolkit~\cite{blei2003latent}, which applies an optimized Gibbs sampling algorithm to assign probabilities to words within documents, generating a topic distribution. This allows us to analyze the distribution of $k$ topics across different financial needs.

To determine the optimal number of topics we are computing skewness of topic distribution of each need instead of perplexity and coherence score to ensure a balance between content strength and distribution~\cite{thukral2022understanding}.

For a given need \( N_i \), let \( T_1, T_2, \ldots, T_k \) be the topic distribution from LDA MALLET. We calculated metrics like mean, standard deviation, median, and skewness for each need, defined as:
\[
\text{Skewness}(N_i) = \frac{3(\mu(N_i) - M(N_i))}{\sigma(N_i)}
\]
where \( \mu(N_i) \) and \( M(N_i) \) are the mean and median, and \( \sigma(N_i) \) is the standard deviation. The goal is to minimize \( W_k \), the count of needs with negative skewness, by iteratively increasing \( k \) until no significant improvement is seen.

\subsection{Behavioral Attributes}

We extracted financial stress and risk propensity to assess users' ability to meet financial needs. Financial stress reflects the feeling of being unable to meet financial demands, with four levels: Low stress, Slightly Stressed, Moderately Stressed, and Very Stressed. Risk propensity indicates willingness to take financial risks, categorized into three levels: (i) cautious, (ii) calculative, and (iii) chance-taking. Emotions from the posts were also extracted using the \texttt{text2emotion} library in Python.

\section{Results}
\label{sec:results}

We first discuss the findings related to needs extraction and their hierarchy, followed by the additional insights we arrive at through topic modelling, and subsequently the insights related to behavioral attributes.

The majority of financial posts and needs come from users in the 21-30, followed by 31-40 age groups. Users in the other age groups make up around 15\% of the sample. There is a strong positive correlation between age and average monthly income up to the 51-60 age group, where income peaks. After that, income declines in the ``more than 60" group to \$6,815.81, lower than all groups except the under-20 group (Table~\ref{tab:user_stats}).
\begin{table}[t]
    \centering
    \caption{Users, Monthly Income, Financial Posts, and Financial Needs by Age Group}
    \begin{tabular}{|l|c|c|c|c|}
    \hline
    age    &  no. of users & avg. monthly income & financial posts & financial needs\\ \hline 
    less than 20    &  17 & 6173.66 & 485 & 1331\\ 
    21-30     & 209 & 6743.91 & 3911 & 10906\\
    31-40     & 66 & 7805.12 & 1488 & 4154\\
    41-50     & 16 & 7094.91 & 278 & 763\\
    51-60     & 13 & 9158.35 & 278 & 714\\
    60 and above     & 13 & 6815.81 & 269 & 733\\
    \hline
    \end{tabular}
    \label{tab:user_stats}
\end{table}


\begin{table}[t]
    \centering
    \caption{Income and Financial Need Distribution by needs hierarchy and prioritization framework}
    \begin{tabular}{|l|c|c|}
    \hline
    Hierarchy Level & Monthly Income & Financial Needs \\ \hline
    Panel A:
NHF & & \\ \hline
Basic Needs & 6536.45 & 4709 \\
Safety Needs & 7189.45 & 11913 \\
Love \& Belongingness & 7568.25 & 134\\
Esteem Needs & 7111.83 & 1764 \\
Self-Actualization & 8410.25 & 81 \\ \hline
Panel B:  NPF & & \\ \hline
Consumption For Immediate Needs & 6774.77 & 2315 \\
Savings For Emergencies & 6952 & 9977 \\
Retirement Savings Wealth  & 7231.73 & 6309 \\ 
and Lifestyle Improvement & & \\\hline 
    \end{tabular}
    \label{tab:users_income_needs}
\end{table}

 In Panel A of Table~\ref{tab:users_income_needs}, we observe that a user’s monthly income increases from \$6536 to \$7568 as we move from basic needs to love and belonging needs. The income for esteem needs drop to \$7111, before rising to \$8410.25 for self-actualization. For the NHF, the mapping of income is slightly weaker due to the reduced income for users in the esteem needs group. 
In Panel B of Table~\ref{tab:users_income_needs}, we observe that users' monthly income increases from \$6,774 to \$6,952 to further to \$7,231 as we move from consumption needs, savings for emergencies, to retirement savings, wealth, and lifestyle improvements for NPF.

We thus observe that hypothesis H1 is supported, except for a slight dip in the average income for the esteem needs.
However, we find full support for hypothesis H2.



\subsection{Topic Extraction Results}
\begin{figure}[t]
    \centering
    \includegraphics[width=0.9\linewidth]{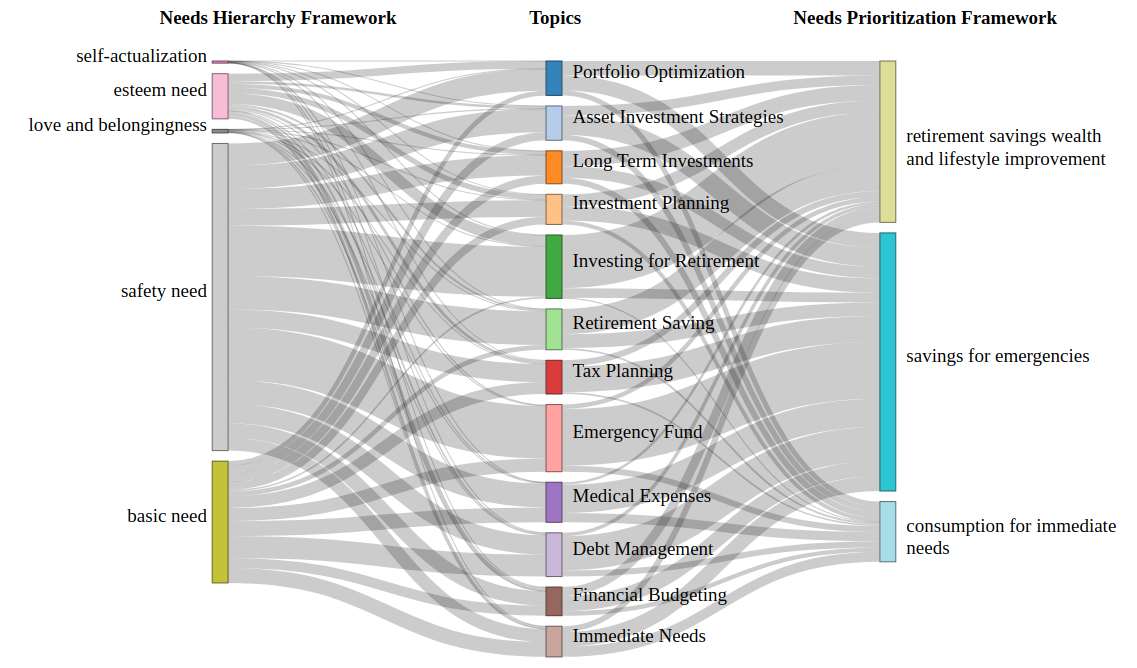}
    \caption{Interconnections Between Financial Topics and Hierarchical Needs Framework}
    \label{fig:topic_map}
\end{figure}

As discussed in Section~\ref{sec:topicex}, 
we extracted 12 optimal topics from our data set of $18,601$ financial needs. We assigned names through each topic’s representative words. These 12 topics are the following -- Investment Planning, Long Term Investments, Portfolio Optimization, Financial Budgeting, Retirement Saving, Immediate Needs, Emergency Fund, Tax Planning, Debt Management, Medical Expenses, Investing for Retirement, Asset Investment Strategies.
We have mapped topics to both NHF and NPF. Thereby we can identify which of the financial aspects are most discussed or prioritized in our dataset.
Figure~\ref{fig:topic_map} depicts how the topics are mapped to the needs hierarchy levels of the two frameworks.

\subsubsection{Mapping of Topics to NHF}
    \textit{Basic Needs}: Emergency Fund (531) and Medical Expenses (598) stand out as critical components, reflecting their necessity-driven nature.
     \textit{Safety Needs}: The highest alignment is observed with Emergency Fund (2118) and Investing for Retirement (2021), reinforcing their role in long-term security. 
    \textit{Esteem Needs}: Portfolio Optimization (307) and Investing for Retirement (441) dominate, linking wealth-building strategies to personal aspirations.
    \textit{Love and Belongingness}: Minimal alignment is noted, with Investing for Retirement and Medical Expenses showing the highest but still low values (19 and 21 respectively), suggesting this need has limited financial expression in social media.
    \textit{Self-Actualization}: Tax Planning (22) and Retirement Saving (11) lead, indicating that advanced financial planning contributes to personal growth. Again, this need is limited in social media.

\subsubsection{Mapping of Topics to NPF}
    \textit{Savings for Emergencies} (9970) is the highest priority across all financial strategies, with Emergency Fund leading (2283 instances), followed by Debt Management (1374) and Medical Expenses (1140).
    \textit{Retirement Savings, Wealth, and Lifestyle Improvement} (6306) is dominant for Investing for Retirement (2144), emphasizing its long-term nature. 
    \textit{Consumption for Immediate Needs} (2315) peaks in Immediate Needs (402), reflecting its short-term focus.

Figure~\ref{fig:topic_map} shows that the topics signify the underlying needs of users, and we also interestingly observe a hierarchy of topics that correlate to a parallel hierarchy of needs. The pyramidal structure propounded by NHF also seems to be applicable to topics, which is more clearly observed when topics are mapped to the NPF.

\subsubsection{Comparing the two Frameworks}
\begin{figure}[t]
    \centering
    \includegraphics[width=0.9\linewidth]{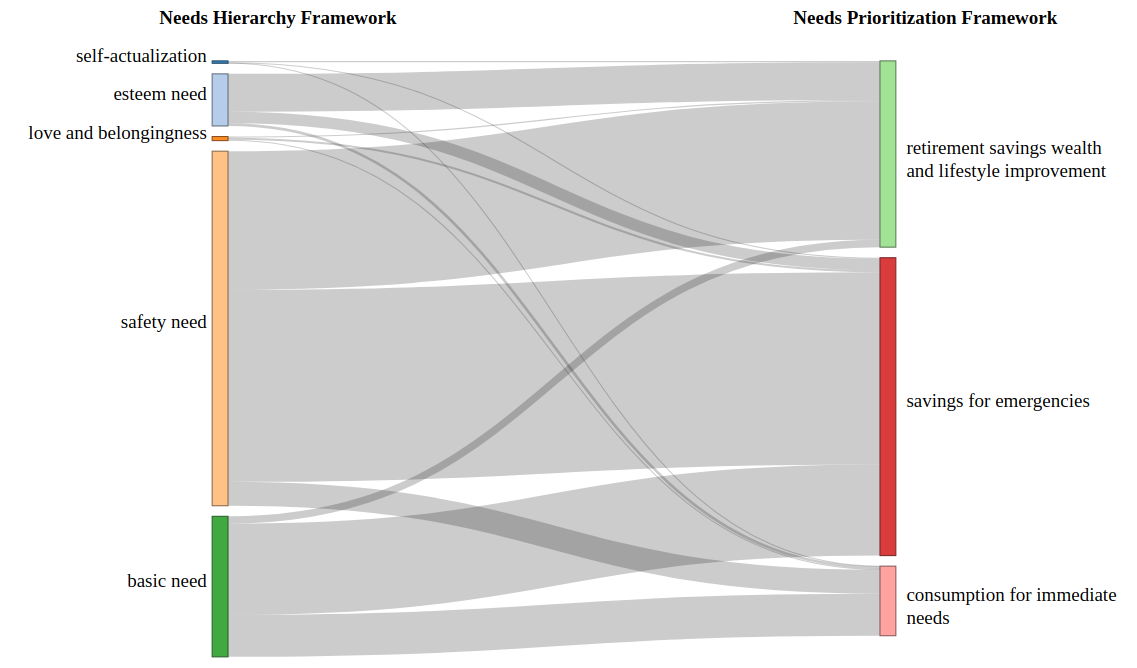}
    \caption{Mapping the Relationship Between Needs Hierarchy and Needs Prioritization}
    \label{fig:hier_map}
\end{figure}

We map the extracted needs to the two frameworks to assess alignment. The mapping (Figure~\ref{fig:hier_map}) shows a strong alignment between immediate consumption, emergency savings, and retirement savings with basic and safety needs from the NHF. The mapping of the levels of the two frameworks is presented below.

``Consumption for Immediate Needs" mainly aligns with basic (1467) and safety needs (833), focusing on survival and security. ``Savings for Emergencies" is strongly tied to safety (6703) and basic needs (3177), highlighting financial preparedness. ``Retirement Savings, Wealth, and Lifestyle Improvement" aligns with safety (4833) and esteem needs (1320), emphasizing long-term stability and lifestyle improvement. Higher-level needs, like love, belonging, and self-actualization, are minimally represented, indicating a priority on survival, security, and future stability over psychological or self-fulfillment goals.

\begin{figure}[t]
\centering
     \includegraphics[width=0.9\linewidth]{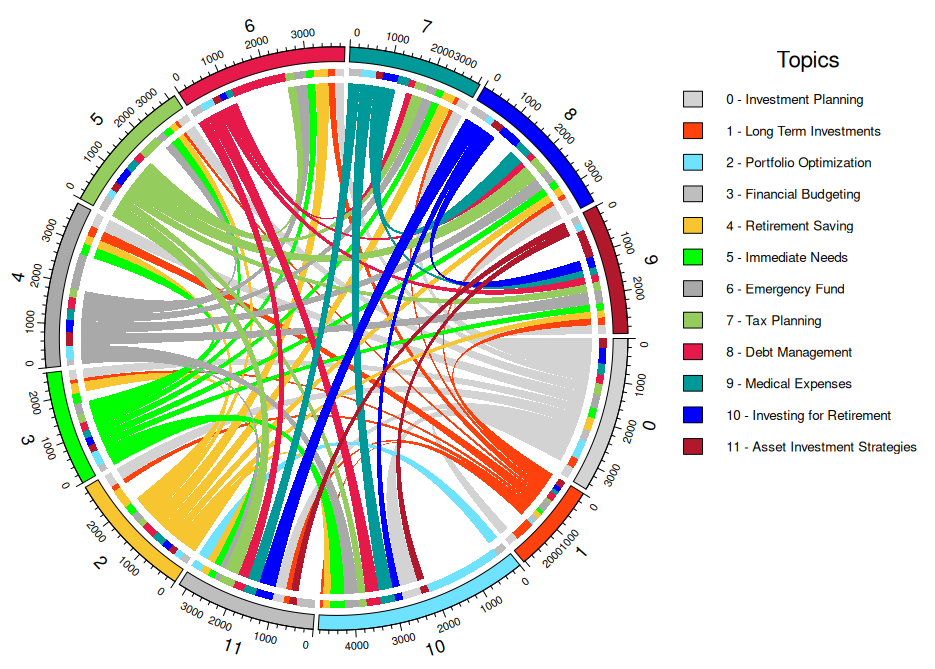}
    \caption{Co-occurrence relationships among financial need topics}
    \label{fig:topic_coocc}
\end{figure}

The chord diagram (Figure~\ref{fig:topic_coocc}) shows a topic co-occurrence graph, highlighting the interconnectivity of financial planning topics. Co-occurrence values indicate how often topics appear together, reflecting their associations in financial decision-making. Investing for Retirement and Emergency Fund are key nodes, emphasizing their central role in financial discussions.

Strong co-occurrences are seen between Investing for Retirement and Retirement Saving (343), highlighting their importance for retirement security. Emergency Fund often co-occurs with Medical Expenses (186) and Immediate Needs (217), emphasizing emergency preparedness. Portfolio Optimization links to Asset Investment Strategies (192) and Tax Planning (303). Debt Management connects with Immediate Needs (410) and Tax Planning (375), reflecting its role in managing liabilities. The link between Tax Planning and Asset Investment Strategies (306) highlights the importance of tax efficiency in investments.
\begin{figure}[t]
    \centering
    \includegraphics[width=\linewidth]{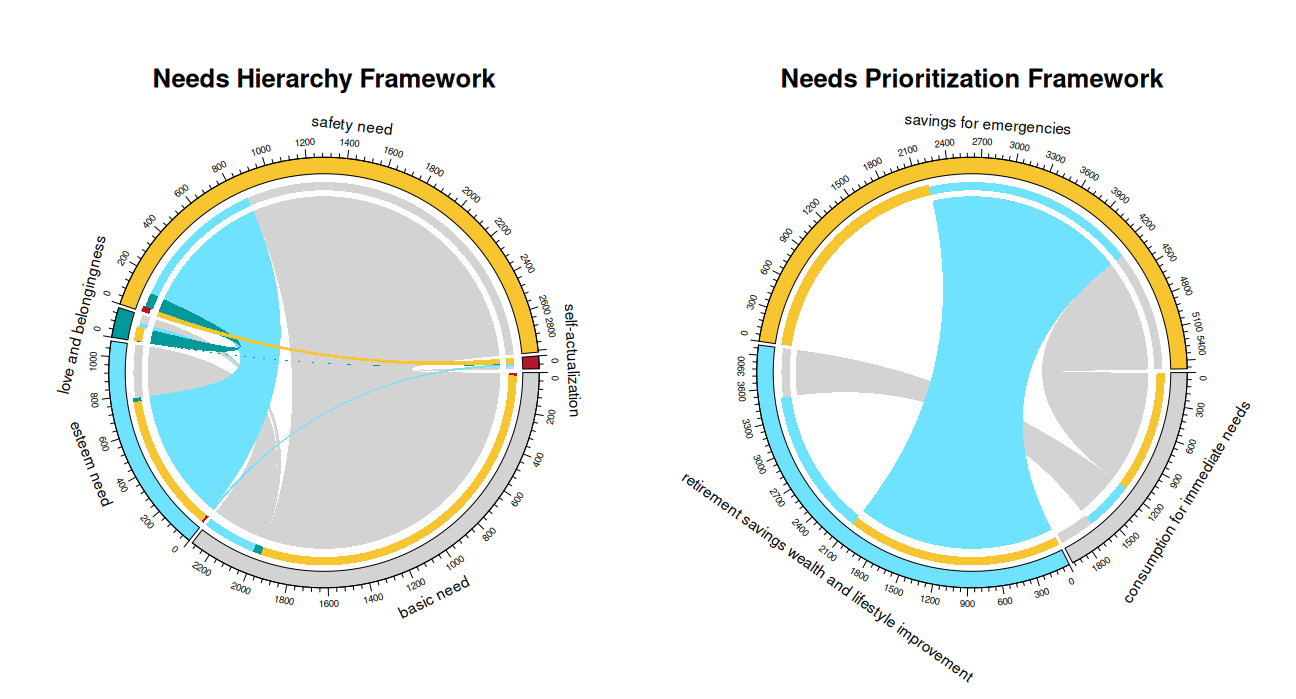}
    \caption{Co-occurrence of Need labels from two Frameworks}
    \label{fig:hier_coocc}
\end{figure}

\subsubsection{Co-occurrences of needs levels for both frameworks}

In the NHF, ``Savings for emergencies" has the highest self-occurrence (2203), highlighting its importance, followed by ``retirement savings, wealth, and lifestyle improvement" (1420). There is a strong link between ``basic needs" and ``safety needs," reflecting their interdependence, in line with Maslow’s framework. ``Esteem needs" (340) show moderate connection to ``safety needs" (751), suggesting safety supports confidence. Higher-level needs like ``love and belonging" (9) and ``self-actualization" (13) are less frequent, indicating they are secondary to foundational needs.

In the NPF, ``Savings for emergencies" has the highest self-occurrence (2203), followed by ``retirement savings, wealth, and lifestyle improvement" (1420). ``Consumption for immediate needs" has the lowest self-co-occurrence (335). The strongest link is between ``savings for emergencies" and ``retirement savings" (2090), highlighting their interdependence. A notable connection exists between ``consumption for immediate needs" and ``savings for emergencies" (1182), but the link with retirement savings is weaker (476), showing limited integration with long-term goals. all the above findings are depicted in Figure~\ref{fig:hier_coocc}.

\subsection{Behavioral Attributes Mapping to Financial Needs}

We analyzed behavioral traits related to stress, risk, and emotions across financial needs, hierarchies, and income levels. Most individuals report moderate stress (41.25\%), indicating manageable financial pressure. Very stressed individuals (11.36\%) face critical challenges, while low (23.39\%) and slightly stressed (24.00\%) levels suggest relative stability. 
Stress distribution by income, needs hierarchy, and financial priorities is detailed in Table~\ref{tab:cat_stress_risk}.
We also compute correlations between Need levels and the behavioral attributes (Table.~\ref{tab:cor_hir_stress}).
For the NHF, correlations show that basic needs are strongly linked to high stress (r = 0.40 with very stressed, r = -0.31 with not stressed), while esteem needs correlate with lower stress (r = 0.25 with not stressed). Safety needs peak under slight stress (r = 0.22) but decline under high stress (r = -0.35). Self-actualization and belongingness show little variation. Overall, stress aligns with lower-level needs; low stress relates to higher-order goals.
For the NPF, high stress correlates with consumption for immediate needs (r = 0.23), while low stress aligns with retirement savings (r = 0.30). Emergency savings rise with stress (r = 0.23) but fall under low stress (r = -0.25), suggesting stress shifts focus from long-term planning to immediate needs.

Risk propensity is dominated by “calculative risk-taking” (70.83\%), reflecting informed decision-making. “Cautious” behavior accounts for 24.69\%, while impulsive types like “chance taking” (3.06\%) and “unassigned” (1.42\%) are rare.
In terms of correlations, basic needs are linked to cautious behavior (r = 0.28) and negatively to calculative (r = -0.22) and chance-taking (r = -0.21), showing risk aversion at lower levels. Esteem needs correlate strongly with chance-taking (r = 0.47), indicating greater risk tolerance. Safety needs align slightly with calculative behavior (r = 0.20). Higher-level needs generally show more risk-taking.
Immediate consumption and emergency savings correlate slightly with caution (r = 0.11) and negatively with chance-taking. In contrast, retirement and lifestyle savings show lower caution (r = -0.26) and higher links with calculative (r = 0.19) and chance-taking (r = 0.28), reflecting greater risk appetite in long-term planning.

For emotions, fear leads (81.22\%), followed by sadness (10.52\%), with surprise, happiness, and anger all minimal. These trends likely influence both stress and risk. See Table~\ref{tab:cat_stress_risk} for detailed distributions.

\begin{table}[t]
\caption{Financial Stress and Risk Distribution Across Income and Needs levels of the two frameworks.}
\resizebox{\textwidth}{!}{
\begin{tabular}{|l|llll|lll|}
\hline
& \multicolumn{4}{c|}{\textbf{Stress levels}}   & \multicolumn{3}{c|}{\textbf{Risk levels}}  \\ \hline
Category                                                                 & \multicolumn{1}{l|}{\begin{tabular}[c]{@{}l@{}}Low\\ stress\end{tabular}} & \multicolumn{1}{l|}{\begin{tabular}[c]{@{}l@{}}Slightly\\ stressed\end{tabular}} & \multicolumn{1}{l|}{\begin{tabular}[c]{@{}l@{}}Moderately\\ stressed\end{tabular}} & \begin{tabular}[c]{@{}l@{}}Highly\\ stressed\end{tabular} & \multicolumn{1}{l|}{Cautious} & \multicolumn{1}{l|}{Calculative} & \begin{tabular}[c]{@{}l@{}}Chance\\ taking\end{tabular} \\ \hline
\textbf{Income}  & \multicolumn{1}{l|}{} & \multicolumn{1}{l|}{}  & \multicolumn{1}{l|}{} &  & \multicolumn{1}{l|}{} & \multicolumn{1}{l|}{} &  \\ \hline

0-4000       & \multicolumn{1}{l|}{1587} & \multicolumn{1}{l|}{1574}  & \multicolumn{1}{l|}{2716} & 912   & \multicolumn{1}{l|}{1901} & \multicolumn{1}{l|}{4580} &  208\\ 

4001-8000       & \multicolumn{1}{l|}{1179}  & \multicolumn{1}{l|}{1247}   & \multicolumn{1}{l|}{2237}   & 580  & \multicolumn{1}{l|}{1178}  & \multicolumn{1}{l|}{3816} &  172 \\ 

8001-12000     & \multicolumn{1}{l|}{784} & \multicolumn{1}{l|}{818}  & \multicolumn{1}{l|}{1301}  & 286 & \multicolumn{1}{l|}{689} & \multicolumn{1}{l|}{2371} &    88    \\ 

12001-16000    & \multicolumn{1}{l|}{388} & \multicolumn{1}{l|}{380}  & \multicolumn{1}{l|}{600}   &  130  & \multicolumn{1}{l|}{390} & \multicolumn{1}{l|}{1049}  &  33 \\ 

16001-20000   & \multicolumn{1}{l|}{158} & \multicolumn{1}{l|}{210}  & \multicolumn{1}{l|}{308}  & 42 & \multicolumn{1}{l|}{171} & \multicolumn{1}{l|}{511}  &  26 \\ 

$>$20000   & \multicolumn{1}{l|}{253}  & \multicolumn{1}{l|}{238} & \multicolumn{1}{l|}{506}  &  167 & \multicolumn{1}{l|}{263}& \multicolumn{1}{l|}{848} &  43 \\ \hline

\textbf{NHF} & \multicolumn{1}{l|}{}          & \multicolumn{1}{l|}{}   & \multicolumn{1}{l|}{}  &  & \multicolumn{1}{l|}{}  & \multicolumn{1}{l|}{}   &   \\ \hline

Basic needs  & \multicolumn{1}{l|}{937}  & \multicolumn{1}{l|}{908}   & \multicolumn{1}{l|}{2143}  & 721 & \multicolumn{1}{l|}{1460} & \multicolumn{1}{l|}{3094} &   48\\ 

Safety Needs    & \multicolumn{1}{l|}{2787} & \multicolumn{1}{l|}{2983} & \multicolumn{1}{l|}{4878}  & 1265 & \multicolumn{1}{l|}{2990} & \multicolumn{1}{l|}{8526}&  304 \\ 

Love \& Belongingness   & \multicolumn{1}{l|}{48}  & \multicolumn{1}{l|}{25}  & \multicolumn{1}{l|}{48}  & 13     & \multicolumn{1}{l|}{19}  & \multicolumn{1}{l|}{91}  &     2 \\ 

Esteem Needs    & \multicolumn{1}{l|}{543}  & \multicolumn{1}{l|}{540}   & \multicolumn{1}{l|}{567} & 114 & \multicolumn{1}{l|}{105}  & \multicolumn{1}{l|}{1415} &     213 \\ 

Self actualization   & \multicolumn{1}{l|}{34} & \multicolumn{1}{l|}{11}  & \multicolumn{1}{l|}{32}  & 4 & \multicolumn{1}{l|}{18} & \multicolumn{1}{l|}{49} & 3  \\ \hline

\textbf{NPF} & \multicolumn{1}{l|}{}          & \multicolumn{1}{l|}{}   & \multicolumn{1}{l|}{}  &  & \multicolumn{1}{l|}{}  & \multicolumn{1}{l|}{}   &   \\ \hline

\begin{tabular}[c]{@{}l@{}}Consumption for\\ immediate needs\end{tabular} & \multicolumn{1}{l|}{527}         & \multicolumn{1}{l|}{451}   & \multicolumn{1}{l|}{1015}  & 322 & \multicolumn{1}{l|}{615} & \multicolumn{1}{l|}{1590}     &  38 \\ 

Savings For Emergencies    & \multicolumn{1}{l|}{2046}  & \multicolumn{1}{l|}{2106}  & \multicolumn{1}{l|}{4375} & 1450 & \multicolumn{1}{l|}{3215} & \multicolumn{1}{l|}{6476}  &  125\\ 

\begin{tabular}[c]{@{}l@{}}Retirement Savings Wealth\\ and Lifestyle Improvement\end{tabular} & \multicolumn{1}{l|}{1776} & \multicolumn{1}{l|}{1910}   & \multicolumn{1}{l|}{2278}  & 345  & \multicolumn{1}{l|}{762}  & \multicolumn{1}{l|}{5109} &   407  \\ \hline
\end{tabular}
}
\label{tab:cat_stress_risk}
\end{table}



\begin{table}[t]
    \centering
    \caption{Correlation between stress levels, risk levels and Need Levels of two frameworks}
    \resizebox{\textwidth}{!}{
    \begin{tabular}{|l|c|c|c|c|c|c|c|}
    \hline
    & \multicolumn{4}{c|}{Stress levels}   & \multicolumn{3}{c|}{Risk levels} \\ \hline
    Category  & Low & Slightly & Moderately & Highly & Cautious & Calculative & Chance \\
    & stress  & stressed & stressed & stressed & & & taking \\\hline 
     \textbf{NHF} &   &  &  & & & & \\ \hline 
    Basic Needs    &  -0.31 & -0.24 & 0.2 & 0.4 & 0.28 & -0.22 & -0.21\\ 
    Safety Needs     & 0.15 & 0.22 & -0.06 & -0.35 & -0.17 & 0.20 & -0.09 \\
    Love \& Belongingness  & -0.04 & -0.02 & 0.02 & 0.04 & 0.08 & -0.08 & 0.01\\
    Esteem Needs     & 0.25 & 0.06 & -0.23 & -0.1 & -0.17 & 0.04 & 0.47 \\
    Self Actualization     & 0.11 & -0.02 & -0.01 & -0.08 & -0.05 & 0.04 & 0.01 \\ \hline
    \textbf{NPF} &   &  &  & & & & \\ \hline
    Consumption for Immediate Needs     & -0.2 & -0.13 & 0.13 & 0.23 & 0.11 & -0.10 & -0.08\\
    Savings for Emergencies     & -0.25 & -0.15 & 0.21 & 0.23 & 0.11 & -0.04 & -0.31 \\
    Retirement Savings Wealth & 0.3 & 0.24 & -0.28 & -0.3 & -0.26 & 0.19 & 0.28 \\ 
   \& Lifestyle Improvement &  &  &  & & & & \\ 
    \hline
    \end{tabular}
    }
    \label{tab:cor_hir_stress}
\end{table}

\section{Discussion}

This paper explores financial needs hierarchies using social media data, where users post their issues and questions for the community to respond to. Our work fills a gap in understanding how these needs evolve. 
By applying Generative AI, we bring empirical support to the hierarchy present in financial needs, thus bridging a gap in literature.
 Our analysis shows moderate stress and calculated risk-taking among users. Key contributions include: (i) a new methodology for studying financial need hierarchies, (ii) using Generative AI to analyze financial needs, demographics, and behaviors from social media, and (iii) comparing these insights with traditional models. This research enhances understanding of financial needs, benefiting both academic research and financial decision-making, helping practitioners forge meaningful consumer interactions for greater business impact.

However, this study does not analyze the temporal aspect of the data. The insights are limited to active social media users, primarily from the US, and may not be generalizable to other cohorts or geographies. We may have to analyze needs in a similar way for users from other geographies as well. A user-centric study will allow insights into individual's needs and financial behavior.

Our analysis supports hypotheses H1 and H2 (Section~\ref{sec:hypo}), showing that financial needs follow a hierarchy and can be mapped to both frameworks (Section~\ref{sec:results}). We also found a positive correlation between income and hierarchy levels (Table~\ref{tab:users_income_needs}). Additionally, techniques like topic modeling reveal the key components of each financial need hierarchy, while behavioral attributes help understand users' urgency in meeting these needs.

The study maps need and income correlations using cross-sectional data, and future research can explore replies, behavioral parameters (e.g., financial awareness, stress markers), and the impact of social discussions. Additionally, insights at the individual level could inform interventions for regulators or advisors. From a practice standpoint, financial professionals can use this understanding to motivate clients to adopt and maintain needs-based financial behaviors.

\bibliographystyle{spmpsci}
\bibliography{asonam}

\end{document}